\documentclass[12pt]{iopart}
\usepackage{graphicx,epsfig}
\usepackage{setstack,iopams,bm}
\begin{document}

\title[Coupling-geometry-induced temperature scales in the conductance of LL wires]{Coupling-geometry-induced temperature scales in the conductance of Luttinger liquid wires}
\author{P.~W\"achter$^1$, V.~Meden$^2$, K.~Sch\"onhammer$^1$}

\address{$^1$ Institut f\"ur Theoretische Physik, Universit\"at G\"ottingen, D-37077 G\"ottingen, Germany} 
\address{$^2$ Institut f\"ur Theoretische Physik A, RWTH Aachen University and JARA - Fundamentals of Future Information Technology, D-52056 Aachen, Germany}
\ead{waechter@theorie.physik.uni-goettingen.de}

\begin{abstract}
We study electronic transport through a one-dimensional, finite-length quantum wire of correlated 
electrons (Luttinger liquid) coupled at arbitrary position via tunnel barriers 
to two semi-infinite, one-dimensional as well as stripe-like (two-dimensional) leads, 
thereby bringing theory closer towards systems resembling setups realized in experiments.
In particular, we compute the temperature dependence of the linear conductance $G$ of a system without bulk 
impurities. The appearance of new temperature scales introduced by the lengths of overhanging 
parts of the leads and the wire implies a $G(T)$ which is much more complex than the power-law 
behavior described so far for end-coupled wires. Depending on the precise setup the 
wide temperature regime of power-law scaling found in the end-coupled case is broken up in up 
to five fairly narrow regimes interrupted by extended crossover regions. Our results can be used 
to optimize the experimental setups designed for a verification of Luttinger liquid 
power-law scaling.
\end{abstract}

\submitto{\JPCM}
\maketitle

\section{Introduction}
\label{intro}

Theoretically, it is well established that the low-energy physics of a wide class of one-dimensional 
(1d) metallic electron systems (quantum wires) with two-particle interaction is 
described by the Luttinger Liquid (LL) phenomenology~\cite{Haldane1}. One of the characterizing properties 
of LL physics is the power-law scaling of a variety of physical observables as function of external parameters.
For spin-rotationally invariant and spinless systems the corresponding exponents can all 
be expressed in terms of a single parameter, the LL-parameter $K$. On the experimental side the situation 
is less clear. Much effort has been put into measurements aiming at a verification of LL behavior by 
observing the predicted power-law scaling. However, in many of those experiments sources for the observed 
behavior other than LL physics cannot be ruled out in a completely satisfying 
way~\cite{KS}.

\begin{figure}[tb]
\begin{center}
\includegraphics[scale=0.5,clip]{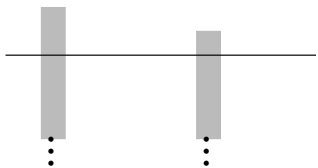}
\end{center}
\caption[]{Sketch of a typical experimental setup. The quantum wire is attached to higher dimensional 
leads, which further extend as indicated by the arrows; note the overhanging parts of the leads 
and the wire.}\label{sketchsys}
\end{figure}

The gap between the status of theory and experiment is partly related 
to an insufficient theoretical modeling of the experimentally available setups. In many 
experiments on quantum wires realized e.g.\ by single wall carbon nanotubes or semiconductor 
heterostructures, the temperature dependence of the linear conductance $G$ was measured~\cite{Exp}. 
A typical experimental system in which the finite-length LL wire is coupled via 
tunnel barriers to two quasi two-dimensional, stripe-like (Fermi liquid) leads is 
sketched in figure \ref{sketchsys}. In the preparation the precise position of the contact 
regions and its width, as well as the quality of the contacts is still difficult to 
control. In contrast to the experimental geometry the theoretical description is 
mostly done by considering end-contacted wires, in which the lead electrons tunnel into the end of the wire and the 
leads terminate at the contacts. Here we partly bridge the gap between the simplicity of the theoretical modeling 
and the complexity of the experimental setup by considering a microscopic model of a quantum wire with 1d as well as stripe-like (two-dimensional) leads, which arbitrarily couple to the wire. Undoubtedly, in order to achieve direct comparison to experimental data true ab-initio simulations of the specific experimental setup would be desirable. Yet, the available ab-initio methods do not capture LL physics and thus cannot be used in the present context. Therefore, microscopic modeling is the method of choice to extend the LL physics originally derived within effective field-theories. We show that the length scales set 
by the overhanging parts of the leads and the wire introduce new energy scales, 
which strongly affect the temperature dependence of the conductance. The $G(T)$ 
curves become much more complex than the ones obtained from the simplified 
modeling. 

Throughout this paper we consider the case of spinless fermions. We expect that 
our results---with the appropriate changes~\cite{KS} in the equations relating scaling 
exponents to the LL parameter $K$---directly carry over to the case of 
spin-$1/2$ electrons if the two-particle interaction is long-ranged (unscreened) 
in real space (that is, if the $g_{1,\perp}$ term of the g-ology 
classification~\cite{Solyom} can be neglected) and spin-rotationally invariant. This is 
generally assumed to be the case in  single wall carbon nanotubes~\cite{SWNTLL}. For semiconductor heterostructures 
the range of the interaction can strongly depend on the screening 
properties of the surrounding environment (free carriers). In the presence of a 
sizable backscattering component ($g_{1,\perp}>0$) the asymptotic low-energy regime 
is usually reached via a complex crossover 
behavior~\cite{MedenHF,Yue,AndergTg0}. We expect the same to hold for the more realistic 
setups studied here, which would render the behavior of $G(T)$ even more complex (and 
less ``universal'') than described in the present work.
       
The most elementary understanding of the temperature dependence of the linear conductance of a LL wire can be gained using Fermi's Golden Rule like arguments. The local 
single-particle spectral function $\rho$ of a translationally invariant, infinite LL
scales as $\rho \sim (\mbox{max}\{ \omega, T \})^{\alpha_{\rm bulk}}$, with $\omega$ 
measured relative to the chemical potential and $\alpha_{\rm bulk}=(K+K^{-1}-2)/2$~\cite{KS}. 
For a semi-infinite LL and $\omega,T \ll v_F/x$, with $x$ being the distance from the 
boundary and $v_F$ denoting the Fermi velocity, the power-law scaling of the now 
$x$-dependent spectral weight is instead given by the boundary exponent 
$\alpha_{\rm end}= K^{-1}-1$~\cite{KF,Fabrizio}. Beyond the energy scale $v_F/x$ the bulk 
exponent is recovered. Fermis Golden Rule then predicts, $G(T) \sim T^{\alpha_{\rm bulk}}$ for tunneling 
into the bulk of a LL while tunneling into the end leads to 
$G(T) \sim T^{\alpha_{\rm end}}$. The LL parameter $0 < K < 1$ (for repulsive interactions; $K=1$ 
in the noninteracting case) is model dependent~\cite{Haldane1,KS}. Only for a few cases (for an 
example see below) its exact dependence on the model parameters such as the strength of the 
two-particle interaction and the filling of the band is known analytically. For small 
interactions parameterized by the amplitude $U$, $K$ generically goes as 
$K=1-U/U_c+{\mathcal O}([U/U_c]^2)$ with the model and parameter dependent scale $U_c$. 
From this Taylor expansion one can infer that $\alpha_{\rm end}$ is linear in $U/U_c$ and for
small interactions dominates over ${\alpha_{\rm bulk}}$ which is quadratic.  

In experimental setups the particles leave the LL wire at a second contact. To obtain the conductance of this type of system one is tempted to simply take the inverse of the sum of the two resistances resulting from the tunnel barriers. 
In the presence of inelastic scattering processes this is certainly correct. 
In the absence of such processes, an approximation used in the present paper, it is however less clear if 
this leads to the correct result for the conductance. In \cite{EnssTrans}  and \cite{Jakobs} 
it was shown using scattering theory that for low-transmittance contacts and at temperatures 
sufficiently larger than the scale $v_F/L$, with the Fermi velocity $v_F$ and the length $L$ of the 
interacting wire, adding the two resistances and taking the invers gives the correct result. We note in passing that 
this can not be extended to the case of more than two impurities~\cite{Jakobs}. 

The Fermi Golden Rule like considerations on the conductance of a LL wire usually do not take 
the geometry of the reservoirs into account explicitely but only use their Fermi liquid properties.
For the case of end-contacted wires the above results were confirmed within a microscopic modeling of the semi-infinite leads 
(1d noninteracting tight-binding chain) and the wire (1d tight-binding chain 
of spinless fermions with nearest-neighbor interaction)~\cite{EnssTrans,Jakobs}. The 
linear conductance shows power-law scaling $T^{\alpha_{\rm end}}$ for temperatures 
$v_F/N \ll T \ll B$, with $N$ the number of lattice sites 
in the interacting wire and the $B$ bandwidth. Other approaches to include the end-contacted leads into 
the model are based on the so-called local Luttinger liquid picture~\cite{LLL} and on the 
concept of radiative boundary conditions~\cite{EggerBC}, where the wire is 
modeled by an effective field-theory (bosonization)~\cite{KS}. 

We here extend the functional renormalization group (fRG) approach used to study
the linear response transport~\cite{EnssTrans} as well as the finite bias steady-state 
nonequilibrium transport~\cite{nonequi} through an end-contacted wire 
(described by a microscopic model) to investigate the 
role of the contacts and the leads more thoroughly. The reservoirs are modeled as two-dimensional 
tight-binding stripes of variable width (including the case of 1d leads) and contacted to the 
1d interacting wire at arbitrary positions. This generically leads to 
overhanging parts of the wire as well as of the leads 
(see figure \ref{sketchsys}). We show that they set new energy scales and the temperature range 
of power-law scaling  of $G(T)$, $v_F/N \ll T \ll B$, breaks up into up to five different regimes of
variable size. This leads to a temperature dependence of $G(T)$ which is much richer than the 
simple scaling $T^{\alpha_{\rm end}}$, which for sufficiently long wires, holds over several decades 
in $T$ within the simplified modeling described above. Our results
show that it is difficult to find clear evidence for LL behavior of $G(T)$ in the generically 
used setups. 

To fully cover all relevant temperature regimes and treat systems of experimental length, i.~e. 
in the micrometer range, we have developed an algorithm which allows us to compute the conductance
for systems of up to $N=10^5$ lattice sites. Including the stripe-like leads requires a 
substantial extension of the order $N$ algorithm developed for 1d leads~\cite{AndergfRG}.

The rest of the paper is organized as follows. In section \ref{Model} the microscopic model used to 
describe the quantum wire and the leads is introduced. In section \ref{scacoI} and the appendix it is shown 
in detail, how to calculate the conductance through the system using scattering theory. Section \ref{fRG} is 
a short survey of the fRG method and the generalizations necessary to include stripe-like 
leads are described. In section \ref{res} our results for the temperature dependence of the conductance are presented 
and discussed. We conclude with a summary in section \ref{summary}.

\section{The Microscopic Model}
\label{Model}

\begin{figure}[tb]
\begin{center}
\includegraphics[width=0.45\textwidth,clip]{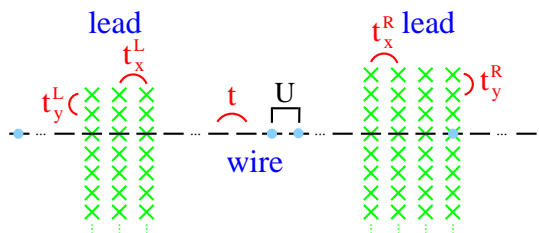}
\end{center}
\caption[]{(Color online) Sketch of the 1d wire connected to stripe-like leads; the crosses 
depict the sites of the leads, the bars those of the wire; note that the coupling between 
the wire and the leads is located at arbitrary positions leading to overhanging parts.}\label{sketchsystem}
\end{figure}

We consider spinless
fermions on a finite 1d lattice with nearest-neighbor hopping $t$ and 
nearest-neighbor two-body interaction $U$, coupled to two semi-infinite, 
stripe-like,  and noninteracting tight-binding leads. For simplicity all lattice constants are assumed to be equal and chosen 
to be unity. The system is sketched in figure \ref{sketchsystem}.
The Hamiltonian $H_{\rm wire}$ of the $N$ site quantum wire reads ($t \geq 0$)
\begin{eqnarray*}
H_{\rm{wire}} = H_{\rm{kin}}+H_{\rm{int}}
\end{eqnarray*}
with 
\begin{eqnarray}
H_{\rm{kin}}&=&-t\sum_{j=1}^{N-1}\left(c^\dagger_jc_{j+1}+\rm{H.c.}\right) \nonumber \\
\label{hamwire}
H_{\rm{int}}&=& U \sum_{j=1}^{N-1} \left(c^\dagger_jc_j-\frac{1}{2}\right)\left(c^\dagger_{j+1}c_{j+1}
-\frac{1}{2}\right),
\end{eqnarray}
where $c^{(\dagger)}_j$ denotes the fermionic annihilation (creation)
operator in Wannier states at site $j\in\{1,...,N\}$.
We here consider the case of half-filling of the leads as well as of the wire. 
To assure the latter the density operator $n_j=c^\dagger_jc_j$ in the last line of (\ref{hamwire}) is shifted by $-1/2$. We expect similar results to hold away from 
half-filling. 

The leads are assumed to be noninteracting
\begin{equation*}
H_{\rm lead}^{a}=-\sum_{\vec{n},\vec{m}\in\mathbf{A}}
\left(t_{\vec{n},\vec{m}}^{a}d^{\dagger}_{\vec{n},a}d_{\vec{m},a} +
 {\rm H.c.}\right)~,
\end{equation*}
where $a$ stands for $L$ (left) and $R$ (right) and the symbol
$\mathbf{A}=\{(n_x,n_y)| n_x\in\{1,...,N_{a,x}\}$ $\wedge n_y\in\{1,...,\infty\}\}$ denotes
 the set of lattice points of lead $a$. The extensions of the leads
 in the $x$-direction given by $N_{a,x} $ are allowed to differ.
The hopping matrix elements $ t_{\vec{n},\vec{m}}^{a} $ are assumed to connect nearest
neighbors only.

The 1d wire and the stripe-like leads are coupled via hopping terms, described by 
\begin{equation}\label{coupl}
H_{\rm coupl}^{a}=\sum_{j=1}^N\sum_{\vec{n}\in\mathbf{A}}
\left(V^{a}_{j,{\vec{n}}}c^\dagger_jd_{\vec{n},a} + {\rm H.c.}\right)~.
\end{equation}
 The matrix elements $V^{a}_{j,{\vec{n}}} $ will be
 specified later.

 The full Hamiltonian $H$ of the system reads
\begin{equation}\label{hamsys}
H=H_{\rm wire}+H_{\rm lead}^{L}+H_{\rm lead}^{R}+
H_{\rm coupl}^{L}+H_{\rm coupl}^{R}\quad.
\end{equation}
The model for an infinite isolated wire $H_{\rm wire}$ with interaction
$U$ between all sites can be solved exactly using the Bethe ansatz~\cite{YangYang}. The
system is a LL for all fillings and interactions except for
half-filling at $|U|>2t$, where a charge density wave forms ($U>2t$)
or phase separation occurs ($U<-2t$). From the Bethe ansatz solution,
 the LL parameter $K$ can be computed and reads for half filling and
   $|U|\leq 2t$~\cite{Haldane}
\begin{equation}\label{exK}
K^{-1}=\frac{2}{\pi}{\rm arcos}\left(-\frac{U}{2t}\right) ~.
\end{equation}

\section{Single particle scattering }
\label{scacoI}
\subsection{General relations}
\label{GenRel}

The fRG procedure in the approximation scheme described in section \ref{fRG} yields a 
frequency independent self-energy of the interacting wire at the end of the fRG flow. 
This can be interpreted as a single-particle (scattering) potential and incorporated 
into a {\it renormalized} single-particle Hamiltonian. As described in detail 
in \cite{EnssTrans} for end-coupled wires with 1d leads one can compute the 
conductance from single-particle scattering theory~\cite{Oguri0} using a generalized 
Landauer-B\"uttiker approach~\cite{LandBuett}.

We therefore consider the single-particle Hamiltonian
\begin{equation}\label{hamonepar}
h = \tilde h_{\rm wire}+h_{\rm lead}^{L}+h_{\rm lead}^{R}
+h_{\rm coupl}^{L}+h_{\rm coupl}^{R}\quad. 
\end{equation}
where $\tilde{h}_{\rm wire}$ is the effective single-particle Hamiltonian
for the wire at the end of the fRG flow (which includes the generated scattering potential) 
and the $h$'s are single-particle versions of the terms introduced in the last section.
For the states in the single-particle Hilbert space we use standard Dirac notation. 
In the scattering problem to be solved, the first three terms of $h$ are considered 
to be the unperturbed part $h_0$, while the couplings to the leads 
$h_{\rm coupl}^{L}+h_{\rm coupl}^{R}$ are considered as the perturbation $h_1$.
We closely follow the steps for the corresponding scattering problem with end-contacted, 1d 
leads~\cite{EnssTrans}.

The main difference of working with higher dimensional leads is the
occurrence of transverse scattering channels. The orthonormal single-particle
eigenstates $|k,l,a\rangle$ of the isolated semi-infinite leads are  
standing waves with  the wavenumber  $k\in [0,\pi]$  in the
$y$-direction and $l=1,...,N_{x,a}$ labeling the transverse modes 
\begin{eqnarray*}
\langle\vec{n},a|k,l,a\rangle &=&
\sqrt{\frac{2}{\pi}}\sin(kn_y)\\
&&\times\sqrt{\frac{2}{N_{x,a}+1}}    \sin\left
  (\frac{l\pi}{N_{x,a}+1}n_x\right) \\
& = & \langle n_y|k\rangle
\langle n_x|l\rangle ^a  = {^a\langle} \vec n|\vec k\rangle^a \; .
\end{eqnarray*}
The corresponding eigenenergies are
\begin{eqnarray}\label{disp2D}
\epsilon_a(k,l)&=&-2t_{y}^a\cos(k)-2t_{x}^a\cos\left(\frac{l\pi}{N_{x,a}+1}
\right  ) \nonumber \\
& = & \epsilon_a(k)+\epsilon_{a,l} = \epsilon_a(\vec k)  ~,
\end{eqnarray}
with the nearest-neighbor hopping matrix elements $t_{x}^a $ and  $t_{y}^a $.

We now consider the scattering of a particle from the left to the
right lead (or vice versa) coupled via $h_1$ to the wire.
The outgoing scattering states
$|k,l,a+\rangle = |\vec k+\rangle^a$ follow from the 
Lippmann-Schwinger equation~\cite{Taylor}
\begin{equation}
|\vec k+\rangle^a=|\vec k \rangle^a+{\mathcal G}(\epsilon_a(\vec k)+i0)h_1|\vec k\rangle^a\,\,,
\end{equation}
where ${\mathcal G}(z)=\left(z-h\right)^{-1}$ is the
resolvent of the full single-particle Hamiltonian.

If we define $\bar a$ as the ``complement'' of $a$ this yields
\begin{equation}
^{\bar a}\langle\vec{m}|\vec k,+\rangle^a
=\sum_{j,\vec n} {^{\bar a}\langle}\vec{m}|{\mathcal G}(\epsilon_a(\vec k)+i0)|j\rangle
 V^{a}_{j,{\vec{n}}}{^a\langle}\vec{n}|\vec k\rangle^{a} 
\end{equation}
 Using ${\mathcal G}={\mathcal G}_0+{\mathcal G}_0 h_1 {\mathcal G}$ with 
${\mathcal G}_0(z)=\left(z-h_0\right)^{-1}$ we obtain 
\begin{equation}\nonumber
^{\bar a}\langle\vec{m}|\vec{k},+\rangle^{a}=
\sum_{j,j'\atop\vec m',\vec n}
  { ^{\bar a} \langle} \vec{m}|{\mathcal G}_0|\vec m' \rangle^{\bar a}
V^{\bar a}_{\vec m',j'}
\langle j'|\mathcal{G}|j\rangle V^{a}_{j,\vec n } {^a\langle}\vec n|\vec k \rangle^{a}.
\end{equation}
Only elastic scattering $k,l,a \to k',l',\bar a$ occurs with
$\epsilon_a(k,l)=\epsilon_{\bar a} (k',l')$. This is
enforced by the free resolvent matrix elements $^{\bar a}\langle \vec m'
|{\mathcal G}_0(\epsilon_a(k,l)+i0)|\vec m \rangle^{\bar a }$.
It is easier to see this in a mixed representation, local in the
$y$-direction and using the transverse channel number $l'$ as the
additional quantum number
\begin{eqnarray}
\label{mixed}
^{\bar a}\langle m_y&,l'|\vec{k},+\rangle^{a}= \nonumber
\\&\sum_{j,j'\atop\vec m',\vec n }{^{\bar a}\langle} m
_y,l'|{\mathcal G}_0|\vec m'\rangle^{\bar a} V^{\bar  a}_{\vec m',j'}
\langle j'|{\mathcal G}|j\rangle V^{a}_{j,\vec{n}}
{^a\langle}\vec{n}|\vec{k}\rangle^{a} \nonumber \\
&=\biggl(-\sum_{j,j'\atop\vec m',\vec n }
\frac{2\pi}{v^{\bar a}(k')}{^{\bar a}\langle}\vec{k'}|\vec m'
\rangle^{\bar a}
V^{\bar a}_{\vec m',j'}\langle j'|{\mathcal G}|j\rangle  V^{a}_{j,\vec{n}}
\nonumber \\
&\phantom{=\biggl(-\sum_{j,j',\vec m',\vec n
  })}\times {^a\langle}\vec{n}|\vec{k}\rangle^{a}\biggr)
\frac{e^{i k'm_y}}{\sqrt{2\pi}}~.
\end{eqnarray}
The second equality holds for $m_y$ larger than the $m_y'$ involved
in the coupling of the lead and the quantum wire. To derive this relation
we used the result for the resolvent of a semi-infinite
1d chain ($n=1,2,...,\infty$) with nearest-neighbor hopping $t$
\begin{eqnarray}
\label{1D}
\langle m|{\mathcal G}_0^{\rm 1d}(\epsilon+i0)|n\rangle&=&
\frac{-i}{v(\epsilon)}\left(           e^{ik(\epsilon) |m-n|} -
e^{ik(\epsilon) (m+n)  } \right ) \nonumber \\
&=& -\frac{2\sin[k(\epsilon)n]}{v[k(\epsilon)]}  e^{ik(\epsilon)m     }~,
\end{eqnarray}
where the second line holds for $m\ge n$.
For $|\epsilon|<2 t$ the wavenumber
$k(\epsilon)\in [0,\pi]$  is the solution  of
$-2t\cos[k(\epsilon)]=\epsilon$ and the velocity $v(k)$ is
given by $2t\sin(k)$. In \cite{EnssTrans} only the result for $n=1$ was
presented. For $|\epsilon|>2t$ the resolvent matrix element decays
exponentially with the distance $|m-n|$ and are thus ``closed'' channels 
in (\ref{mixed}). This happens if $|\epsilon_a(k,l)-\epsilon_{\bar a,l'}|<2|t_y^{\bar a}
|$, and $k'\in (0,\pi)$ is determined by energy conservation
$\epsilon_{\bar a}(k')=\epsilon_a(k)+\epsilon_{a,l}-\epsilon_{\bar a,l'} $.
The velocity in (\ref{mixed}) is given by
$v^{\bar a}(k')=2t_y^{\bar a}\sin(k')$.
The prefactor of the outgoing plane wave in (\ref{mixed})
gives the transmission probability for the scattering 
$k,l,a \to k',l',\bar a$  
\begin{eqnarray}\nonumber
|t&_{|\vec{k}\rangle^a\rightarrow|\vec{k'}\rangle^{\bar a}}|^2 =
|t(\epsilon_a(\vec k),l,l')|^2\\\label{transampl}
&=\Big|\sum_{j,j'\atop\vec m,\vec n}
\frac{2\pi}{v^{\bar a}(k')}{^{\bar a}\langle}\vec{k'}|\vec{m}
\rangle^{\bar a} V^{\bar a}_{{\vec{m},j'}}
\langle j'|{\mathcal G}|j\rangle V^{a}_{j,{\vec{n}}}
{^a\langle}\vec{n}|\vec{k}\rangle^{a}\Big|^2\,\,.
\end{eqnarray}

Taking the sum over $l$ compatible with $\epsilon_a(\vec k)=\epsilon$ 
and over the $l'$ values of the corresponding ``open'' channels we 
define~\cite{LandBuett}
\begin{equation}
\mathcal{T}_a(\epsilon)=\sum_{\{l,l'\}} 
\frac{v^{\bar a}(\epsilon,l')}{v^a(\epsilon,l)}|t(\epsilon,l,l')|^2~.
\end{equation}
The unitarity of the $S$-matrix guarantees
$\mathcal{T}_a(\epsilon)=\mathcal{T}_{\bar a}(\epsilon) =
\mathcal{T}(\epsilon) $.
This quantity determines the conductance $G(T)$ in the generalized Landauer-B\"uttiker
formula~\cite{LandBuett}
\begin{equation}\label{LandBuett}
G(T)=\frac{e^2}{h}\int {\rm d}\epsilon
\left(-\frac{\partial f}{\partial\epsilon}\right)\mathcal{T}(\epsilon)\,\,,
\end{equation}
where $f$ denotes the Fermi function
and $e^2/h$ is the unit of the quantized conductance. Note that for our interacting 
problem the flowing self-energy and thus $\tilde h_{\rm wire}$ as well as 
the effective transmission probability $\mathcal{T}(\epsilon)$ become 
$T$ and $N$ dependent (for details, see below). 

The Landauer-B\"uttiker formula (\ref{LandBuett}) holds in the abscence of inelastic processes 
due to the two-particle interaction (that is for vanishing imaginary part of the self-energy)~\cite{Oguri0}. 
Such processes do not appear in our approximate treatment of the interaction and using (\ref{LandBuett}) does thus not present an additional approximation.

For the calculation of the transmission probability one has to compute the matrix elements
$\langle j'|{\mathcal G}|j\rangle$ of the full resolvent. This 
can be reduced to the inversion of a $N\times N$-matrix, using 
Feshbach projection~\cite{EnssTrans}
\begin{eqnarray}\nonumber
P\mathcal{G}(z)P&=\\ \label{profo}
&\left[zP-PhP-PhQ(zQ-QhQ)^{-1}QhP\right]^{-1}, 
\end{eqnarray}
where $P$ and $Q=1-P$ are projection operators.
Here we use as $P$ the projection operator $P_{\rm wire}=\sum_{j=1}^N|j\rangle\langle j|$ 
onto the wire and $Q$ the one on the leads.

\begin{figure}[tb]
\begin{center}
\includegraphics[width=0.47\textwidth]{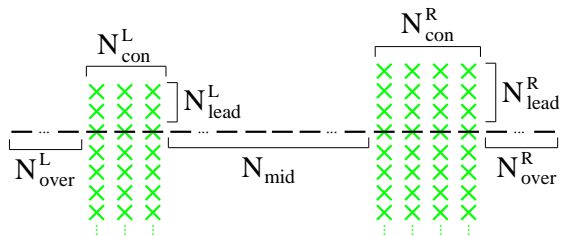}
\end{center}
\caption[]{(Color online) Sketch of the system considered. The overlapping sites of wire and leads are individually coupled via a hopping term. The length scales are indicated.} \label{geolength}
\end{figure}

The special geometry (mimicking a generic experimental setup) used in section \ref{res} is shown in figure \ref{geolength}.
Each site of the wire lying on top of a site of the lead (as shown in the figure) is coupled to this site via a hopping term $t^{c_a}_n$. The length scales appearing in the system are also defined in the figure. Note that the overall length $N=N_{\rm mid}+N_{\rm con}^{L}+N_{\rm con}^{R}+N_{\rm over}^{L}+N_{\rm over}^{R}$.

The projected inverse Green function takes the form shown in figure \ref{strucgreen} of the appendix.
We need the tridiagonal elements of this Green function to calculate the right-hand side of 
the flow equations (\ref{floweq}) and (\ref{floweq2}), see section \ref{fRG}. An algorithm to compute 
these elements is presented in the appendix. Furthermore, to calculate the transmission 
amplitude (\ref{transampl}) we need the $j,j'$-elements of the Green function 
with $j'\in\{N_1+1,..,N_2\}$ and $j\in\{N_3+1,..,N_4\}$, i.~e. the elements connecting the 
left and right contact region. An algorithm to efficiently compute these elements is also briefly 
sketched in the appendix~\cite{waechter}.

For simplicity, we assume the hoppings in the leads and in the wire to be equal and define this 
as our unit of energy, i.~e. $t^{a}_x=t^{a}_y=t=1$. 

\subsection{One-dimensional leads}
\label{scacoII}

The appearance of new temperature scales is most easily seen in the
case of {\it one-dimensional} leads, i.e. $N^L_{\rm con}
=N^R_{\rm con}=1 $. For 
$N\gg N_{\rm lead}^{a}$ and
leads weakly coupled to the wire, 
i.e.~$(t^{c_a})^2\ll 1$ the transmission probability ${\mathcal
  T}(\epsilon) $ consists of 
$N$ narrow resonances near the 
eigenvalues $\epsilon_\alpha$ of the wire without contacts.
 The integrated weight $w_\alpha$ of the Lorentzian resonance near
$\epsilon_\alpha$ is~\cite{EnssTrans}
\begin{equation}\nonumber
w_\alpha\approx 4\pi\left [  \frac{1}{\Delta^L_\alpha} +
  \frac{1}{\Delta^R_\alpha} \right]^{-1}~.
\end{equation}
Here the partial widths $\Delta^a_\alpha $ are given by
\begin{eqnarray*}
\Delta^a_\alpha=(t^{c_a})^2&\frac{2}{(N+1)\sin[k(\epsilon_\alpha )] }\\
&\times\sin^2[N^ak(\epsilon_\alpha )]
\sin^2[(N^a_{\rm lead}+1)k(\epsilon_\alpha)]~,
\end{eqnarray*}
with $N^a$ the site of the wire to which lead $a$ couples.
In the temperature regime  $v_F/N \ll T \ll B$ the Landauer
B\"uttiker formula simplifies to 
\begin{equation}
\label{Gbig}
G(T)\approx\frac{e^2}{h}\sum_{\alpha=1}^Nw_\alpha
\left(-\frac{\partial f}
{\partial\epsilon}\right)_{\epsilon=\epsilon_\alpha}
\end{equation}
 We return to this expression when 
discussing our numerical results. 

\section{The Functional Renormalization Group}
\label{fRG}

In this section we present a short outline of the fRG at finite temperature $T$, which we 
use to treat the interacting system. General aspects of the fRG can 
be found in \cite{Salmhofer1},~\cite{Salmhofer}, and~\cite{MedenVorl}, while its application 
to inhomogeneous LLs is described in \cite{AndergTg0},~\cite{EnssTrans} and~\cite{AndergfRG}.
By comparison to results for small systems obtained by density-matrix renormalization group and 
to exact results from bosonization and Bethe ansatz for the asymptotic behavior, it was shown that 
the fRG in the truncation scheme used here captures the relevant physics not only in the asymptotic 
low-energy regime but also on finite energy scales~\cite{EnssTrans,AndergfRG}.
Towards the end of this section, we describe the extensions of the method necessary when dealing 
with stripe-like leads.

Starting point of our implementation of the fRG is the introduction of an infrared energy cutoff 
$\Lambda$ in the Matsubara frequencies of the noninteracting propagator. This leads to 
a $\Lambda$-dependence of the generating functional of the one-particle irreducible Green 
functions (vertex functions). Taking the derivative of this functional with respect 
to $\Lambda$ leads to an exact, but infinite, hierarchy of coupled first order 
differential equations for the vertex functions. We truncate this hierarchy by neglecting the $m$-particle 
vertices with $m \geq 3$. In order to reduce the numerical effort, we project the 2-particle vertex function onto the Fermi 
points and parameterize it by a nearest-neighbor interaction 
of amplitude $U^\Lambda$~\cite{EnssTrans,AndergfRG}. Finally, we neglect the feedback of 
the 1-particle vertex (the self-energy) onto the flow of the 2-particle vertex function, 
that is the flowing effective interaction~\cite{AndergfRG}. The resulting set of equations is 
integrated from the initial cutoff $\Lambda=\infty$ down to $\Lambda=0$, at which the original,
cutoff-free problem is recovered. For end-contacted, interacting wires and weak to 
intermediate interactions this approximate approach was shown to provide reliable 
results by comparison with exact analytical and numerical results~\cite{AndergTg0,EnssTrans,AndergfRG}. 

Within our approximation the self-energy $\Sigma$ becomes real, frequency-independent and tridiagonal in position space. 
It acquires a dependence on the temperature and size of 
the interacting wire and can thus be interpreted as a ($T$- and $N$-dependent) 
single-particle scattering potential. This leads to an effective single-particle 
Hamiltonian with renormalized onsite energies and hopping matrix elements.

As mentioned in the introduction to leading order the scaling exponent of 
the tunneling conductance into an end-coupled LL is linear in $U$. In contrast, tunneling
into a translationally invariant (bulk) LL is characterized by an exponent which 
for small $U$ is quadratic~\cite{KS}. First order scaling exponents were also 
found in other setups of inhomogeneous LLs (e.g.\ LLs with a single 
impurity \cite{KF,EnssTrans,AndergfRG} and Y-junctions of LLs; see e.g. Ref. \cite{Barnabe} 
). In all cases in which predictions for the scaling exponents of inhomogeneous LLs 
from alternative approaches (e.g.\ effective field-theories) exist those 
were reproduced to linear order using our truncated fRG 
method~\cite{EnssTrans,AndergfRG,Barnabe}. However, our truncation procedure 
does {\it not} capture bulk LL physics, as terms of order $U^2$ are only partly 
kept in the flow equation for the self-energy. In particular, the terms of order 
$U^2 \ln{(\omega)}$ obtained in a perturbative calculation of the self-energy of 
a translationally invariant LL (evaluated at the Fermi momentum $k_F$), leading 
to the bulk LL power laws when resummed by an RG method, are neglected in our 
approach. We thus expect that, in situations which are dominated by bulk LL 
physics (for examples see below) the scaling exponent vanishes within our 
approximation.

The set of flow equations to be solved 
reads~\cite{AndergfRG}
\begin{eqnarray}\label{floweq}
\frac{d}{d\Lambda}\Sigma^\Lambda_{j,j}&=&-\frac{1}{\pi}\sum_{r=\pm 1}U^\Lambda_{j,j+r} \, 
{\rm Re} \, \left[ 
\mathcal{G}^\Lambda_{j+r,j+r}(i\Lambda)\right] \,,\\\label{floweq2}
\frac{d}{d\Lambda}\Sigma^\Lambda_{j,j\pm 1}&=&\frac{1}{\pi}U^\Lambda_{j,j\pm 1} \, 
{\rm Re} \, \left[ \mathcal{G}^\Lambda_{j,j\pm 1}(i\Lambda) \right] \, ,\\\label{floweq3}
U^\Lambda&=&\frac{U}{1+\left(\Lambda-\frac{2+\Lambda^2}{\sqrt{4+\Lambda^2}}\right)
\frac{U}{2\pi}}\quad,
\end{eqnarray}
where we specialized the flow equation for the effective interaction 
to the case of half-filling. The implementation of finite temperatures requires 
at each step of the flow the replacement of the flow parameter $\Lambda$ by 
the nearest Matsubara frequency~\cite{AndergTg0}.

In order to compute the right-hand side of the flow equations for $\Sigma^\Lambda$ one has to compute the 
tridiagonal elements of the matrix $\mathcal{G}^\Lambda(i\Lambda)$ inverting 
the given $N \times N$-matrix $\mathcal{G}^\Lambda(i\Lambda)^{-1}$. 
To cover all relevant energy scales and to treat systems of lengths in the micrometer range we need 
an algorithm, which allows us to calculate these matrix elements more efficiently than 
the standard order $N^3$ methods~\cite{Schwarz}. In the case of 1d leads, 
$\mathcal{G}^\Lambda(i\Lambda)^{-1}$ itself is tridiagonal and the tridiagonal elements can be 
computed in order $N$ time~\cite{AndergfRG,TriInv}. For stripe-like leads 
$\mathcal{G}^\Lambda(i\Lambda)^{-1}$ consists of tridiagonal blocks, connected on the sub- and 
superdiagonal to full blocks, one for each contact region. The explicit structure is depicted 
in figure \ref{strucgreen} of the appendix. There we present an algorithm, in which 
the problem of inverting the full $N \times N$-matrix is split up into the multiple
inversion of each isolated block. Thus, we can readily use the 
order $N$ algorithm for the tridiagonal parts and some standard 
algorithm~\cite{Schwarz} for the full matrices. The bottleneck of the method is therefore 
the width of the stripe-like leads, which determines the size of the full matrices. 

\section{Results}
\label{res}

\subsection{One-dimensional leads arbitrarily coupled to a noninteracting wire}

We first consider the conductance $G(T)$ through a clean 
quantum wire with two semi-infinite 1d leads (stripe-like lead with a width of one lattice site) 
coupled at arbitrary position of the leads 
and the wire. This introduces four additional length scales, namely 
$N_{\rm lead}^{L/R}$ and $N_{\rm over}^{L/R}$, 
apart from the overall length $N$ of the wire (see figure \ref{geolength}). They
strongly effect the temperature dependence of the conductance.

First we shortly comment on even-odd-effects arising from the fact that the coupling 
sites in the wire as well as those in the leads can each be even or odd and so 
can be the overall length of the chain. Within the notation introduced in 
section \ref{GenRel}, the coupling sites in the wire are $N_{\rm over}^{L}+1$ and $N_{\rm over}^{R}+1$, those in the leads 
$N_{\rm lead}^{L}+1$ and $N_{\rm lead}^{R}+1$. 
For $T \rightarrow 0$, the derivative of the Fermi function in (\ref{LandBuett}) 
becomes a $\delta$-function and the integral is given by the effective 
transmission evaluated at the Fermi energy. This shows even-odd-effects. 
For $T \to 0$ the conductance reaches a nonvanishing value only if the coupling sites 
are all \emph{odd}. For even sites the local spectral density at the Fermi 
energy vanishes, leading to $G(T=0) = 0$. If, for odd coupling sites and $t^{c_L}=t^{c_R}$, the overall
length $N$ is taken to be odd, $G(T\rightarrow0)\rightarrow e^2/h$ holds. For temperatures of the order of $v_F/N$ and higher these even-odd effects vanish. We mainly consider systems with odd $N$ and coupling at odd sites. This implies $G(T=0) = e^2/h$ for $t^{c_L}=t^{c_R}$, which also holds for the interacting system.

\begin{figure}[tb]
\begin{center}
\includegraphics[width=0.45\textwidth]{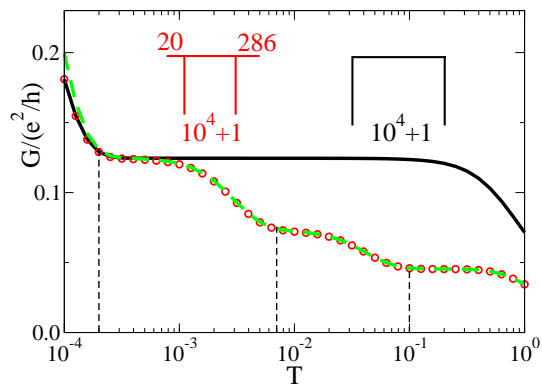}
\end{center}
\caption[]{(Color online) Conductance $G$ as function of temperature $T$ of a 
noninteracting wire of length $N=10^4+1$ and coupling $t^{c_R}=t^{c_L}=0.25$. 
Solid line: end-contacted wire, i.e.~$N_{\rm over}^{L/R}=0$ (setup shown as the right inset) 
obtained from (\ref{LandBuett}). Circles: $N_{\rm over}^{L}=20, N_{\rm over}^{R}=286$ 
(setup shown as the left inset) obtained from (\ref{LandBuett}). Dashed line: the same parameters 
as for the circles but using the approximate analytical result (\ref{Gbig}). The vertical lines 
indicate the crossover scales resulting from the overhanging parts.}\label{1DWW0lengthscales}
\end{figure}

In figure \ref{1DWW0lengthscales}, the conductance $G(T)$ of a noninteracting wire coupled at 
the ends is compared to the conductance of a system coupled in the bulk. The leads are weakly 
connected with $t^{c_R}=t^{c_L}=0.25$ and terminate at the contacts (no overhanging parts, $N_{\rm lead}^{L}=
N_{\rm lead}^{R}=0$). 
For the bulk-coupled case we chose $N_{\rm over}^{L}=20$, 
$N_{\rm over}^{R}=286$ (circles). The conductance of the end-coupled wire (solid line) 
starts at $e^2/h$ for $T = 0$, crosses over into a plateau at a scale $T \approx v_F/N$ (left vertical dashed line) and finally falls off as $T^{-1}$ for $T \gg B$ (band effect). For the bulk-coupled 
case the plateau region breaks up into three clearly distinguishable regimes with the two 
additional crossover scales $v_F/N_{\rm over}^{R}$ and $v_F/N_{\rm over}^{L}$ 
indicated by the other two vertical dashed lines.  
In the first region  with $v_F/N \ll T \ll v_F/N_{\rm over}^{R}$ a plateau with the 
same conductance as for the end-coupled chain appears.  At the two additional scales the 
conductance crosses over to  new plateaus with lower conductance values and finally falls off 
for $T \gg B$. The dashed line shows the conductance of the bulk-coupled 
system using the approximate analytical expression (\ref{Gbig}) for comparison. For temperatures 
in the range $v_F/N \ll T \ll B$ we find excellent agreement with the exact result obtained from 
(\ref{LandBuett}). 

Note that the overhanging parts of the wire and of the leads 
enter symmetrically in (\ref{Gbig}). Thus, choosing $N_{\rm lead}^{L}=20$, 
$N_{\rm lead}^{R}=286$ and $N_{\rm over}^{L}
=N_{\rm over}^{R}=0$, gives an identical conductance as that shown by the 
circles and dashed line in figure \ref{1DWW0lengthscales}. Obviously this will no longer 
hold in the presence of interactions in the wire. 

\begin{figure}[tb]
\begin{center}
\includegraphics[width=0.45\textwidth]{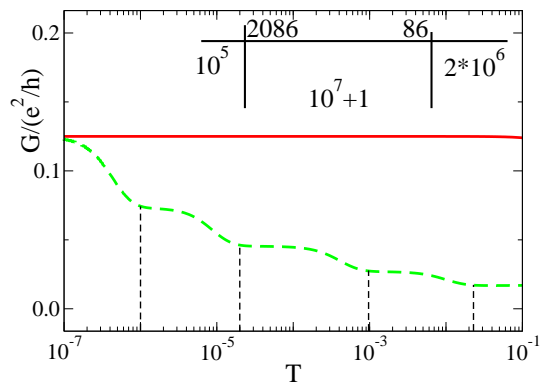}
\end{center}
\caption[]{(Color online) Conductance $G$ of a 
noninteracting wire of length $N=10^7+1$ and coupling $t^{c_R}=t^{c_L}=0.25$. Solid line: end-coupled wire, 
i.e.~$N_{\rm lead}^{L/R}=N_{\rm over}^{L/R}=0$ obtained from the analytical 
result (\ref{Gbig}). Dashed line: $N_{\rm over}^{L}=10^5, 
N_{\rm over}^{R}=2 \cdot 10^6, N_{\rm lead}^{R}=2086$ and 
$N_{\rm lead}^{R}=86$ (setup shown as the inset) obtained from the analytical 
result (\ref{Gbig}). The vertical lines indicate the crossover scales.}\label{hugesys}
\end{figure}

All four new temperature scales, i.e.~$v_F/N_{\rm lead}^{L/R}$ and 
$v_F/N_{\rm over}^{L/R}$  can be resolved using the analytical expression 
(\ref{Gbig}) for a very large system (that is for a very small lower bound $v_F/N$). 
This is shown as the dashed line in figure \ref{hugesys}. The solid line shows the 
conductance for a system coupled at the ends for comparison. 

After identifying the relevance of the additional energy scales due to the 
overhanging parts of the wire and the leads for noninteracting wires we turn
to the interacting case.

\subsection{One-dimensional leads arbitrarily coupled to an interacting wire} 

As described in section \ref{fRG} the approximate fRG flow leads to a nontrivial static self-energy 
for the interacting wire which is tridiagonal in the site indices. At half-filling and for tunnel 
barriers as used here the diagonal part vanishes and the effective potential is given by 
a nonlocal term, the modulated hopping $\Sigma_{j,j+1}$. If the leads are coupled to the ends 
of the wire (with and without overhanging parts of the leads), the two contact 
barriers are the source of (nonlocal) potentials which oscillate 
as $(-1)^{\tilde j}$ (for half-filling) and decay as $1/\tilde j$, where $\tilde j$ denotes 
the distance from the ends (at $j=1$ and $j=N$, respectively). The prefactor of the 
decay is {\it universal} in the sense that it does not depend on the strength of the tunnel 
barriers $t^{c_R}$ and $t^{c_L}$ and is the same even for  $t^{c_R} = t^{c_L} =0$, that is a 
decoupled wire with open boundaries. The power-law decay is cut off at a scale $j_{T} \propto 1/T$ 
set by the temperature beyond which $\Sigma_{j,j+1}$ decays exponentially.
Scattering off this effective potential leads to the power-law scaling of the local spectral weight 
and the conductance discussed in the introduction with the exponents determined by the universal 
prefactor~\cite{Yue,EnssTrans}.  

\begin{figure}[tb]
\begin{center}
\includegraphics[width=0.45\textwidth]{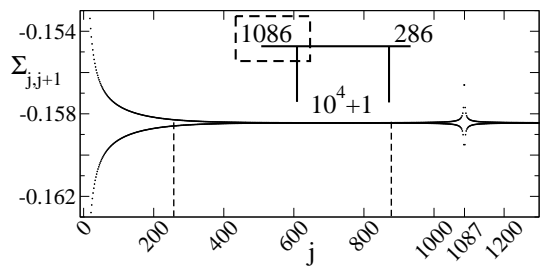}
\end{center}
\caption[]{Off-diagonal self-energy at the end of the fRG flow for an interacting wire 
with $U=0.5$ and couplings $t^{c_R}=t^{c_L}=0.25$ at $T=4 \cdot 10^{-3}$. The wire of length $N=10^4+1$ is coupled to the end of the leads (i.e.~$N_{\rm lead}^{L}=N_{\rm lead}^{R}=0$) with $N_{\rm over}^{L}= 1086$, $N_{\rm over}^{R}=286$.
The part shown in the main plot is indicated by the dashed box in the inset. Long-ranged oscillations emerge from the boundaries and from the coupling site. The scale $j_T$ beyond which the oscillations are damped exponentially is indicated by the 
vertical dashed lines.}\label{selferg}
\end{figure}

A typical off-diagonal component of the 
self-energy $\Sigma_{j,j+1}$ of a bulk-coupled interacting wire is 
shown in figure \ref{selferg}. From the ends of the wire the $(-1)^{\tilde j}/\tilde j$ oscillations 
with universal prefactor emerge and become exponentially damped at a distance
$j_{T} \propto 1/T$ (indicated by the vertical dashed line). For clarity only one 
end is shown in the main plot. The coupling to the lead (at site 1087) introduces similar 
oscillations but with a {\it nonuniversal} prefactor. In contrast to the situation 
found close to an end-contact these oscillations do {\it not} lead to a power-law suppression of 
the spectral  weight close to the coupling 
site resulting from the local inhomogeneity. Furthermore, as discussed in section \ref{fRG} 
our truncated fRG procedure does not capture {\it bulk} LL power laws with exponents 
of order $U^2$. Thus, within our approximation the spectral weight close to 
the coupling site in the bulk shows {\it no} power-law behavior at all.

\begin{figure}[tb]
\begin{center}
\includegraphics[width=0.45\textwidth]{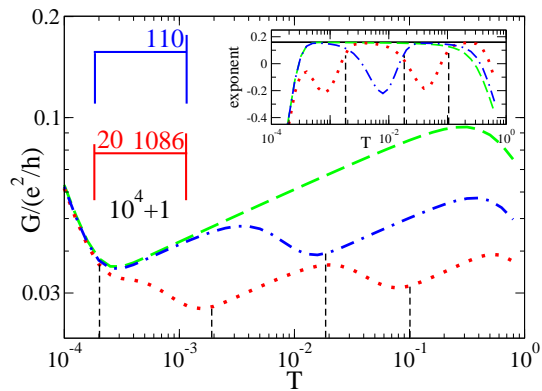}
\end{center}
\caption[]{(Color online) Main plot: Conductance $G$ of an interacting wire with $U=0.5$ of length $N=10^4+1$ 
as function of the temperature $T$. The coupling to the leads is located at the end of the 
wire ($N_{\rm over}^{L}= N_{\rm over}^{R}=0$) and has 
a small amplitude $t^{c_R}=t^{c_L}=0.25$. Dashed line: $N_{\rm lead}^{L/R}=0$. Dashed-dotted line: 
$N_{\rm lead}^{L}=0, N_{\rm lead}^{R}=110$.
Dotted line: $N_{\rm lead}^{L}=20, 
N_{\rm lead}^{R}=1086$. The vertical dashed lines terminating at the different curves 
indicate the corresponding crossover scales.  Left inset: Setups studied. Right inset: Effective exponents. 
The solid horizontal line indicates the fRG approximation for $\alpha_{\rm end}$ as obtained in \cite{EnssTrans}.}\label{WWp5Rand}
\end{figure}

\begin{figure}[hb]
\begin{center}
\includegraphics[width=0.45\textwidth]{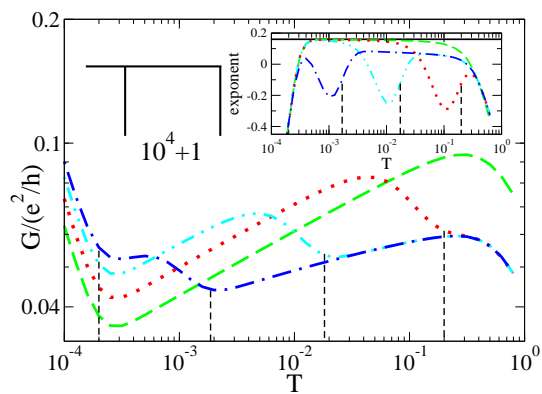}
\end{center}
\caption[]{(Color online) As in figure \ref{WWp5Rand} but with $N_{\rm lead}^{L} = 
N_{\rm lead}^{R}=0$, $N_{\rm over}^{R}=0$ and for 
different $N_{\rm over}^{L}$. Dashed line:  $N_{\rm over}^{L}=0$. 
Dotted line: $N_{\rm over}^{L}=11$. Dashed-double-dotted line: 
$N_{\rm over}^{L}=111$. Dashed-dotted line: $N_{\rm over}^{L}=1111$.}\label{fromleft}
\end{figure}

\begin{figure}[hb]
\begin{center}
\includegraphics[width=0.45\textwidth]{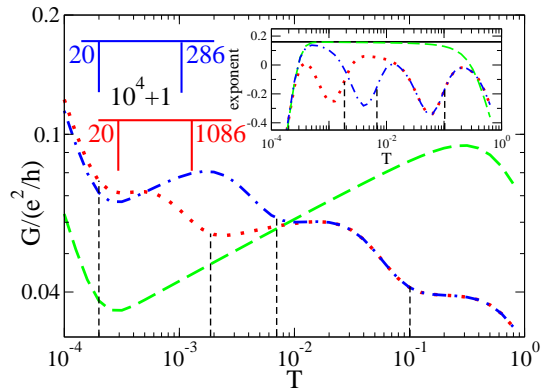}
\end{center}
\caption[]{(Color online) The same as in figure \ref{WWp5Rand} but with $N_{\rm lead}^{L}= N_{\rm lead}^{R}=0$ and coupling into the bulk. Dashed line: $N_{\rm over}^{L/R}=0$. Dashed-dotted line: 
$N_{\rm over}^{L}=20, N_{\rm over}^{R}=286$ 
Dotted line: $N_{\rm over}^{L}=20, N_{\rm over}^{R}=1086$.
}\label{WWp5Bulk}
\end{figure}

Figures \ref{WWp5Rand} to \ref{WWp5Bulk} show the conductance 
for several parameter sets and temperatures 
larger than $v_F/N$, i.e.~temperatures in the regime in which in the noninteracting case the 
plateaus emerge. The length of the wire is $N=10^4+1$ sites which corresponds to 
roughly a micrometer taking typical lattice constants. 
The vertical dashed lines terminating at the different curves indicate the new 
energy scales for the corresponding parameter set.
The curves for the corresponding 
end-coupled systems without any 
overhanging parts are included in the figures for comparison (dashed lines). The 
effective exponent of $G(T)$, computed as the logarithmic derivative of the conductance, 
is shown as the right inset in each figure. The left inset shows the setups of interest.

In figure \ref{WWp5Rand} we focus on wires with $N_{\rm over}^{L}= 
N_{\rm over}^{R}=0$. For all three temperature regimes, 
\begin{eqnarray*}
& v_F/N \ll T \ll v_F/{\rm max}(N_{\rm lead}^{L/R}), \\ 
& v_F/{\rm max}(N_{\rm lead}^{L/R}) \ll T \ll v_F/{\rm min}(N_{\rm lead}^{L/R})\quad {\rm and}\\
& v_F/{\rm min}(N_{\rm lead}^{L/R})\ll T \ll B,
\end{eqnarray*}
the usual power law $G(T)\sim T^{\alpha_{\rm end}}$ is found. Note that for the system with 
$N_{\rm lead}^{L}=20$, $N_{\rm lead}^{R}=1086$ (dotted line) 
the first temperature regime $v_F/N\ll T \ll v_F/N_{\rm lead}^{R}$ is too 
small for a plateau in the effective exponent to develop. However, 
by reducing $N_{\rm lead}^{R}$ one can extend this regime, such 
that the effective exponent tends towards $\alpha_{\rm end}$.

Already this simple case with no overhanging parts of the wire (that is the particles 
tunnel into the end of the interacting wire) demonstrates our central 
result. The power law, which for end-contacted wires of experimental length without any 
overhanging parts of the leads was found to hold over several decades, is 
split up into significantly smaller temperature regimes (of roughly one decade) 
with power-law scaling interrupted by extended crossover regimes. Thus the three 
plateaus of different conductance obtained in the noninteracting case develop 
into power laws with the same exponent in the presence of two-particle interactions.    
For each temperature regime the appearance of the latter 
can be traced back to the spatial dependence of the self-energy $\Sigma_{j,j+1}$. 
Over the whole temperature range $v_F/N\ll T\ll B$, when moving from the left 
to the right contact the fermions have to pass the long ranged oscillations with 
universal amplitude originating from  the ends of the wires. For $v_F/N\ll T$ these 
are well separated, such that the simple arguments presented previously hold, implying 
that $G(T)\sim T^{\alpha_{\rm end}}$. 

The role of the length scales resulting from overhanging parts is further 
exemplified in figure \ref{fromleft} for systems with $N_{\rm lead}^{L} = 
N_{\rm lead}^{R}=0$ and $N_{\rm over}^{R}=0$ but
varying  $N_{\rm over}^{L}$.
Beginning with a purely end-coupled system ($N_{\rm over}^{L}=0$; dashed line) 
the left coupling site is moved into the bulk of the wire. For temperatures 
$T \ll v_F/N_{\rm over}^{L}$ the particles pass both oscillatory 
potentials originating from the ends of the wire and the situation of fermions 
tunneling in and out of the two ends is effectively recovered. 
This leads to power-law scaling of the conductance with exponent $\alpha_{\rm end}$. 
With increasing $N_{\rm over}^{L}$ the crossover 
scale $v_F/N_{\rm over}^{L}$ decreases and the effective exponent reaches 
$\alpha_{\rm end}$ inside a smaller and smaller temperature regime (see the inset). 
For the system with the largest $N_{\rm over}^{L}$ 
($N_{\rm over}^{L}=1111$; dashed-dotted line in figure \ref{fromleft}), the 
interval $[v_F/N,v_F/N_{\rm over}^{L}]$ becomes too small for the 
logarithmic derivative to reach its asymptotic value (see the inset). 
For temperatures $T\gtrapprox v_F/N_{\rm over}^{L}$ 
the oscillations originating at the left end of the wire are exponentially damped at the position 
of the left contact and to the right of it. On their direct path from the left to the right 
contact the fermions thus only experience the potential originating from the right end of 
the  wire. As discussed in the introduction one such potential is sufficient for the spectral 
weight close to its origin to scale as $\omega^{\alpha_{\rm end}}$. One is thus tempted to 
conclude that $G(T)$ follows a power law with exponent $\alpha_{\rm end}$ 
also for  $v_F/N_{\rm over}^{L}\ll T \ll B$. Figure \ref{fromleft} shows that 
this is not the case. Instead, the effective exponent seems to approach an asymptotic value  
which is significantly smaller than $\alpha_{\rm end}$ (see the dashed-dotted line). The naive 
expectation ignores the complexity of the scattering problem. For example, the fermions can also move to the left 
of the left contact before reaching the right lead. Below we further characterize the behavior in this 
regime.

As a third example in figure \ref{WWp5Bulk} we show results obtained for 
$N_{\rm lead}^{L} = N_{\rm lead}^{R}=0$ (no overhanging parts 
of the leads) and different positions of the coupling to the wire $N_{\rm over}^{L/R}$. 
In the regime $v_F/N \ll T \ll v_F/{\rm max}(N_{\rm over}^{L/R})$, the 
conductance follows the power law $G(T)\sim T^{\alpha_{\rm end}}$ since the 
oscillations of $\Sigma_{j,j+1}$ from the two boundaries reach beyond the positions of the 
contacts. For $N_{\rm over}^{R}=1086$ (dotted line) this 
regime becomes too small for the power law to develop. For $v_F/{\rm max}(N_{\rm over}^{L/R}) \ll T \ll 
v_F/{\rm min}(N_{\rm over}^{L/R})$, the oscillations from the boundary, 
originating at the site of the longer overhanging part, are already cut off as the exponential 
damping sets in at $j_T\propto 1/T$, i.e.~$j_T \ll 
{\rm max}(N_{\rm over}^{L/R})$. In this regime the 
curves seem to show power-law behavior with an exponent smaller than $\alpha_{\rm end}$ similar to the 
behavior discussed in connection with figure \ref{fromleft}. In the third regime 
$v_F/{\rm min}(N_{\rm over}^{L/R})\ll T \ll B$, the oscillations from the boundaries do not reach the region between the left and right contact (due to the exponential damping) and no power-law behavior can be found (see the conductance 
plateau of the dotted and dashed-dotted line in the main part of figure \ref{WWp5Bulk}). This is in accordance with 
the fact that the oscillations of $\Sigma_{j,j+1}$ originating from the bulk contacts 
do not have the same effect on the spectral weight as the ones coming from the boundaries of the wire 
(no power-law suppression). We expect the conductance in this temperature regime to show power-law scaling 
with the bulk LL exponent $\alpha_{\rm bulk} \sim U^2$ not captured by our truncated fRG procedure.  

\begin{figure}[hb]
\begin{center}
\includegraphics[width=0.45\textwidth]{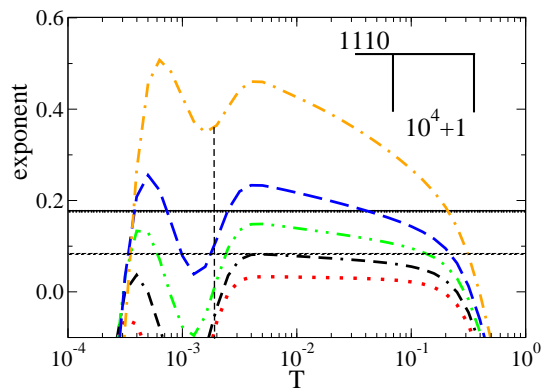}
\end{center}

\caption[]{(Color online) System as in figure \ref{fromleft} with $N_{\rm over}^L=1110$. Effective exponents in the regime $v_F/N_{\rm over}^{L}\ll T\ll B$ for different interaction strengths $U$. From bottom to top: $U=0.25;0.5;0.75;1.0;1.5$. The horizontal lines show the fRG approximation for $\alpha_{\rm end}(U=0.5)/2$ and $\alpha_{\rm end}(U=1.0)/2$ as obtained in \cite{EnssTrans}. The vertical line indicates the crossover scale.}\label{alphaBhalb}
\end{figure}

To obtain a better understanding of the apparent second exponent smaller than $\alpha_{\rm end}$ 
found above, figure \ref{alphaBhalb} shows the logarithmic derivative of $G(T)$ in 
the temperature regime $v_F/{\rm max}N_{\rm over}^{L} \ll T\ll B$ for a system with 
$N_{\rm over}^{L}=1110$ and $N_{\rm over}^{R}=0$. 
Curves for different interaction strengths $U$ are plotted. For $U\leq0.5$ 
(double-dashed-dotted line and below) one is tempted to conclude that $G$ scales like 
$T^{\alpha_{\rm end}/2}$. The lower horizontal line shows $\alpha_{\rm end}/2$ for $U=0.5$ numerically 
calculated within the fRG-implementation used here~\cite{EnssTrans}. However, 
for $U>0.5$ the plateau of the logarithmic derivative becomes strongly 
tilted, such that no definite statement about any power law can be made (compare 
the solid horizontal line showing $\alpha_{\rm end}/2$ for $U=1$ with the dashed curve). Attempts 
to treat the underlying scattering problem, e.g.~with phase-averaging~\cite{Jakobs} or methods developed 
for Y-junctions~\cite{Barnabe} have not yet led to a consistent picture. Furthermore, in this 
temperature regime $G(T)$ might be altered if bulk LL behavior is properly introduced.

The conductance through ``mixed'' systems, that is those with overhanging parts of the quantum wire as well 
as overhanging parts of the leads, shows an even richer temperature dependence. As in the noninteracting 
case (see figure \ref{hugesys}) up to {\it five} different temperature regimes and the 
corresponding crossover regimes might appear in\\ $[v_F/N,B]$.  The behavior 
sufficiently away from the characteristic energy scales set by the lengths of the 
overhanging parts can be understood as explained above. 

These considerations show that for a generic experimental system with four 
overhanging parts all having different lengths and not all of them being several orders 
of magnitude smaller than the total length of the wire it will be very difficult (if not impossible) 
to observe clear indications of a single power law in $G(T)$. We next show that considering 
stripe-like leads, as being realized in the experiments, does {\it not} improve this 
situation.

\subsection{Stripe-like leads arbitrarily coupled to a quantum wire}

We here extend our analysis to stripe-like leads, as realized in many experiments. For this 
setup the parameter space grows linearly with increasing the number of 
coupling sites, e.g.~for five sites at each contact, 
we can freely choose ten couplings $t^{c_{L/R}}_j$. To prevent a proliferation of parameters 
we therefore focus on equal couplings, $t^{c_{L/R}}_j=t^{c_{L/R}}_{\phantom{j}}$.

\begin{figure}[tb]
\begin{center}
\includegraphics[width=0.45\textwidth]{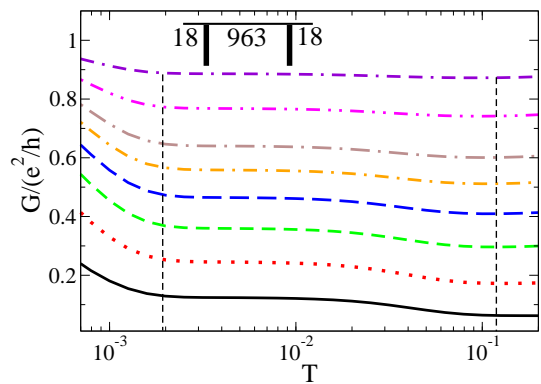}
\end{center}
\caption[]{(Color online) Conductance $G$ as function of temperature $T$ of a noninteracting wire with stripe-like leads of different width. All wire-lead couplings are set equal: $t^{c_L}=t^{c_R}=0.25$. 
The setup is shown in the inset: $N_{\rm over}^L=N_{\rm over}^R=18$, 
$N_{\rm lead}^{L}=N_{\rm lead}^R=0$ and $N_{\rm mid}=963$. From bottom to 
top, the width of the contact region is $N_{\rm con}^L=N_{\rm con}^R=1,3,5,7,9,11,15,21$. 
The vertical lines indicate the crossover scales.} \label{WW0tounit}
\end{figure}

Figure \ref{WW0tounit} shows the conductance of a noninteracting quantum wire with 
$N_{\rm mid}=963$, $N_{\rm over}^L=N_{\rm over}^R=18$, 
$N_{\rm lead}^L=N_{\rm lead}^R=0$ and for varying width of the leads 
$N_{\rm con}^{L/R}$. The vertical lines indicate the temperature scales 
$v_F/N_{\rm over}^{L/R}$ and $v_F/N$, which turned out to be relevant for 
1d leads. They seem to be of equal importance for stripe-like leads 
(see below). The overall shape of the curves is unaltered by increasing
the width of the leads. For large numbers of lead channels $G$ approaches 
the unitary conductance $e^2/h$ of a single channel wire. This trend can
be counteracted by reducing the strength of the lead-wire coupling. 

In the case of 1d leads we identified two pronounced even-odd-effects. The dependence on the total length of the wire 
is also found for stripe-like leads. The vanishing local spectral density at even sites of 
the wire gives rise to the second even-odd-effect. For stripe-like leads 
one coupling site of the wire has to be odd and this effect vanishes as the leads become 
broader.

\begin{figure}[tb]
\begin{center}
\includegraphics[width=0.45\textwidth]{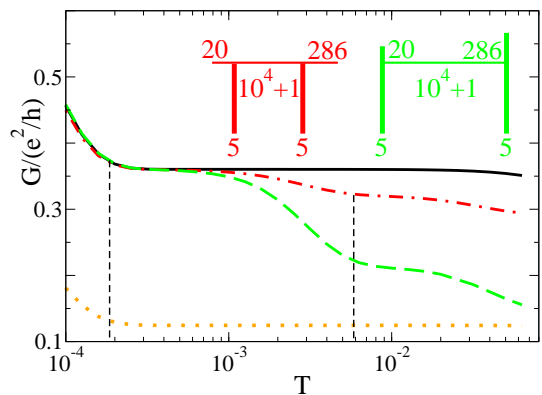}
\end{center}
\caption[]{(Color online) As in figure \ref{WW0tounit} but with length $N_{\rm mid}=10^4+1$ and width $N_{\rm con}^L=N_{\rm con}^R=5$ of contact region.
Dashed-dotted line: $N_{\rm lead}^L=0$, $N_{\rm lead}^R=0$ and $N_{\rm over}^L=20$, $N_{\rm over}^R=286$ (setup shown in left inset). 
Dashed line: $N_{\rm lead}^L=20$, $N_{\rm lead}^R=286$ and $N_{\rm over}^L=N_{\rm over}^R=0$ (setup shown in right inset).
Solid line: $N_{\rm lead}^L=N_{\rm lead}^R=0$ and $N_{\rm over}^L=N_{\rm over}^R=0$.
Dotted line: $N_{\rm con}^L=N_{\rm con}^R=1$, $N_{\rm lead}^L=N_{\rm lead}^R=0$ and $N_{\rm over}^L=N_{\rm over}^R=0$ for comparison. The vertical lines indicate the crossover scales.} \label{2DWW0divL}
\end{figure}

Next we further investigate the relevance of the energy scales identified for 1d leads starting with the noninteracting case. Figure \ref{2DWW0divL} shows the conductance for similar 
configurations as in figure \ref{1DWW0lengthscales}, but with $N_{\rm con}^{L/R}=5$. For 
comparison, the results for the end-coupled system with $N_{\rm lead}^L=N_{\rm lead}^R=0$, 
$N_{\rm over}^L=N_{\rm over}^R=0$, and $N_{\rm con}^L=N_{\rm con}^R=5$ are 
included. Overall, the conductance is considerably higher than in figure \ref{1DWW0lengthscales} as expected 
from the results shown in figure \ref{WW0tounit}. At the temperature scales $v_F/N$, $v_F/N_{\rm over/lead}^L$, and 
$v_F/N_{\rm over/lead}^R$, the conductance subsequently crosses over to lower plateaus. However, 
the complete equivalence of $N_{\rm over/lead}^L$ and $N_{\rm over/lead}^R$ is lost 
even for the non-interacting case as the geometric equivalence is no longer present.

\begin{figure}[tb]
\begin{center}
\includegraphics[width=0.45\textwidth]{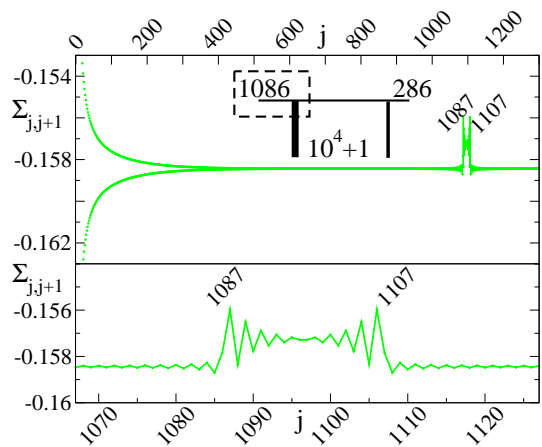}
\end{center}
\caption[]{(Color online) Off-diagonal self-energy at the end of the fRG flow for an interacting wire 
with $U=0.5$ at $T=4 \cdot 10^{-3}$. All wire-lead couplings are set equal: $t^{c_R}=t^{c_L}=0.25$. The setup is shown in the inset: $N_{\rm con}^L=5$, $N_{\rm con}^R=1$, $N_{\rm mid}=10^4+1$, $N_{\rm lead}^{L}=N_{\rm lead}^{R}=0$ and $N_{\rm over}^{L}= 1086$, $N_{\rm over}^{R}=286$. The part shown in the upper plot is indicated by the dashed box in the inset. Long-ranged oscillations emerge from the boundaries and from all coupling sites. The lower plot is a magnification of the left contact region.}\label{SE2D}
\end{figure}

Figure \ref{SE2D} shows a typical off-diagonal component of the self-energy for the case of stripe-like 
leads coupled to the bulk of an interacting wire. For comparison to the case of 1d leads the 
parameters are chosen the same as in figure \ref{selferg}. The oscillations emerging from the 
boundaries are expectedly unaffected by the width of the leads. The contact region itself 
(of size $N_{\rm con}^L$; lower panel in the figure) contains a superposition of 
oscillations emerging from {\it all} coupling sites. The symmetric decay of these oscillations results from the fact that we chose all couplings to be equal. The oscillations originating from the left 
contact to the left and the right contact to the right however still fall off as the inverse distance 
up to a scale $j_{T} \propto 1/T$ beyond which they decay exponentially. As for 1d leads 
these oscillations do not induce  a power-law suppression of the spectral density within 
our fRG approximation. Because of these similarities the temperature dependence of $G$ 
for an interacting wire with stripe-like leads can be traced back to the case of 1d leads. 

\begin{figure}[tb]
\begin{center}
\includegraphics[width=0.45\textwidth]{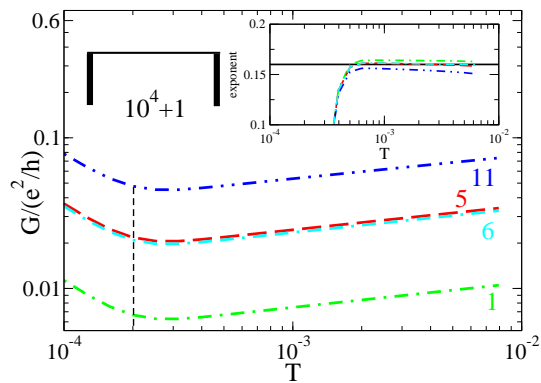}
\end{center}
\caption[]{(Color online) Main plot: Conductance $G$ of an interacting wire with interaction strength $U=0.5$ as function of the temperature $T$ coupled to stripe-like leads of different width. The couplings to the leads are set equal: $t^{c_R}=t^{c_L}=0.1$. The setup is shown in the left inset: $N_{\rm mid}=10^4+1$, $N_{\rm lead}^L=N_{\rm lead}^R=0$, $N_{\rm over}^L=N_{\rm over}^R=0$ 
and $N_{\rm con}^L=N_{\rm con}^R=1;6;5;11$ (from bottom to top as indicated in the plot).
The vertical dashed lines indicate the crossover scales. Right inset: Effective exponents. 
The solid horizontal line indicates the fRG approximation for $\alpha_{\rm end}$ as obtained in 
\cite{EnssTrans}. Line styles as in main part.} \label{2DmWWRanddivZul}
\end{figure}

\begin{figure}[tb]
\begin{center}
\includegraphics[width=0.45\textwidth]{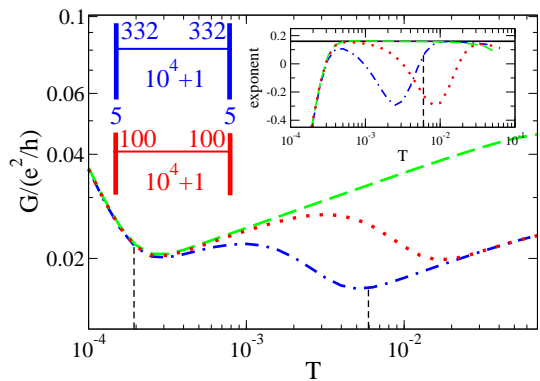}
\end{center}
\caption[]{(Color online) As in figure \ref{2DmWWRanddivZul} but with different $N_{\rm lead}^{L}=N_{\rm lead}^{R}$.
Dashed line: $N_{\rm lead}^{L}=N_{\rm lead}^{R}=0$. Dotted line: $N_{\rm lead}^{L}=N_{\rm lead}^{R}=100$.
Dashed-dotted line: $N_{\rm lead}^{L}=N_{\rm lead}^{R}=332$.}\label{2DWWp5RAND}
\end{figure}

In figure \ref{2DmWWRanddivZul} we focus on end-coupled systems (no overhanging parts)
with $t^{c_L}=t^{c_R}=0.1$ (weak coupling). The only energy scales emerging are $v_F/N$ and $B$. 
For $v_F/N \ll T \ll B$, the conductance scales as $\sim T^{  \alpha_{\rm end}}$. The closeness of the results for $N^{L/R}_{\rm con}=5$ and $6$ shows the minor importance of the even-odd effect.
Figure \ref{2DWWp5RAND} shows systems with tunneling into the end of the wire but with overhanging leads. For comparison the end-coupled case without overhanging 
leads is shown as the dashed line. As in the case of 1d leads the universally decaying 
oscillations from the ends of the system induce power laws $G(T) \sim T^{\alpha_{\rm end}}$ in 
the three temperature regimes introduced by the overhanging parts. For a wire of roughly 
a micrometer in length it is difficult to clearly resolve these power laws due to the extended 
crossover regions and the upper (bandwidth $B$) and lower (cutoff scale $v_F/N$) bounds 
of any power-law scaling. We thus choose $N_{\rm lead}^L=N_{\rm lead}^R$ in 
order to reduce the number of regimes to two. Even in this case the two power laws cannot be resolved 
for a single parameter set (compare the dashed-dotted and dotted lines).

\begin{figure}[tb]
\begin{center}
\includegraphics[width=0.45\textwidth]{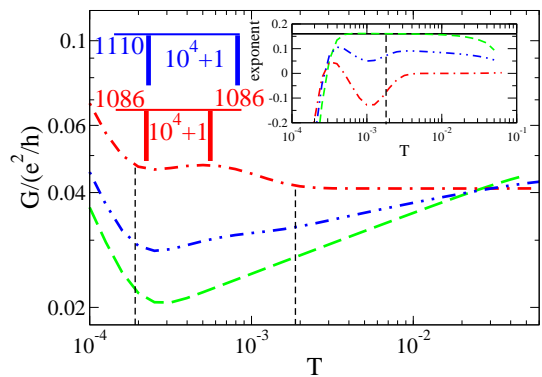}
\end{center}
\caption[]{(Color online) As in figure \ref{2DmWWRanddivZul} but with different $N_{\rm over}^{L}=N_{\rm over}^{R}$.
Dashed line: $N_{\rm over}^{L}=N_{\rm over}^{R}=0$. Dashed-dotted line: $N_{\rm over}^{L}=N_{\rm over}^{R}=1086$.
Dash-dot-dotted line: $N_{\rm over}^{L}=1110, N_{\rm over}^{R}=0$.}\label{2DmWWBULK}
\end{figure}

The case of leads coupled to the bulk of the interacting wire and terminating at the contacts 
is shown in figure \ref{2DmWWBULK}. As anticipated from the noninteracting case 
(figure \ref{2DWW0divL}), the temperature scales introduced by the additional length scales $N_{\rm over}^L$ and 
$N_{\rm over}^R$ are still present. In all three temperature regimes 
\begin{eqnarray*}
& v_F/N \ll T \ll v_F/{\rm max}(N_{\rm over}^{L/R}),\\
& v_F/{\rm max}(N_{\rm over}^{L/R})\ll T \ll v_F/{\rm min}(N_{\rm over}^{L/R})\quad {\rm and}\\
& v_F/{\rm min}(N_{\rm over}^{L/R})\ll T \ll B, 
\end{eqnarray*}
the conductance $G(T)$ is governed by the same behavior as in the case of 1d leads. In the 
first regime the oscillations from the boundaries strongly superpose the coupling sites, thus we find 
$G(T)\sim T^{\alpha_{\rm end}}$. Note that to better resolve the second and third regime, this first regime is
chosen such small in the curves shown that the asymptotic value of the exponent is not 
reached. The second regime is exemplified by the dashed-double-dotted curve. After crossing over at 
scale $v_F/N_{\rm over}^{L/R}$, the conductance \emph{seems} to follow a power law. However, 
as explained in detail in the section on 1d leads, this can not definitely be confirmed and the 
behavior might be altered in the presence of the bulk LL power law. 
In the third regime, the oscillations in the self-energy from the boundaries are cut off by the exponential 
fall-off at scale $\propto 1/T$ before reaching the contact region, thus, no power-law behavior 
can be found within the truncated fRG (figure 16, dashed-dotted line). We expect the conductance in this regime to show scaling 
with the bulk LL exponent if computed with an improved approximation which properly describes also bulk LL behavior.

The conductance of ``mixed systems'' with overhanging parts of the leads and the wire again shows an 
even richer temperature dependence which can be understood by combining the cases studied above. 

\section{Summary}
\label{summary}

In the present study we were aiming at a more realistic modeling of transport through 
interacting 1d quantum wires showing LL physics by considering {\it stripe-like leads} of 
variable width coupled to the wire at {\it arbitrary positions} via tunnel barriers. 
We showed that the overhanging parts of the leads and the wire, which generically 
appear in experimental setups designed for a verification of LL power-law scaling, strongly 
modify the temperature dependence of the linear conductance.  For end-contacted wires of 
experimental length with leads terminating at the contacts the range of possible power-law 
scaling (with exponent $\alpha_{\rm end}$) bound by $v_F/N$ and $B$ extends over roughly four 
decades. In the presence of up to 
four additional energy scales induced by the lengths of the four overhanging parts and 
the corresponding extended crossover regimes the regions of possible scaling 
become so small that the power laws are not visible. This shows the
difficulties to clearly confirm power-law behavior in transport
experiments with LL wires. 

One might speculate that the situation becomes even worse if inelastic two-particle scattering processes neglected 
in our approximate treatment of the interaction (and also in bosonization) are taken 
into account. They set a new upper energy scale $B^\ast$ for the simple scaling behavior considered 
here which is presumably smaller than the band width $B$. 
Power laws might further be obscured by the presence of relaxation and dissipation in the leads, not included in our calculations.

The optimal setup to observe the boundary exponent $\alpha_{\rm end}$ is an end-contacted, 
long wire with no overhanging parts of wire or leads for which one can expect scaling 
to hold for $v_F/N \ll T \ll B^\ast$. Although our approximation does not include the bulk 
LL exponent  the results indicate that the best setup to observe $\alpha_{\rm bulk}$ is 
a long wire with leads which both couple to the bulk of the wire and which 
terminate at the contacts.    

Our approach allows for more general systems
with e.g.~leads of asymmetric widths, random wire-lead couplings or screened interactions in the contact 
regions.

\ack
The authors are grateful to the Deutsche Forschungsgemeinschaft 
(SFB 602) for support.

\begin{appendix}
\section*{Appendix}
\setcounter{section}{1}

\section*{Calculation of resolvent matrix elements}

\begin{figure} 
\begin{center}
\includegraphics[width=0.35\textwidth]{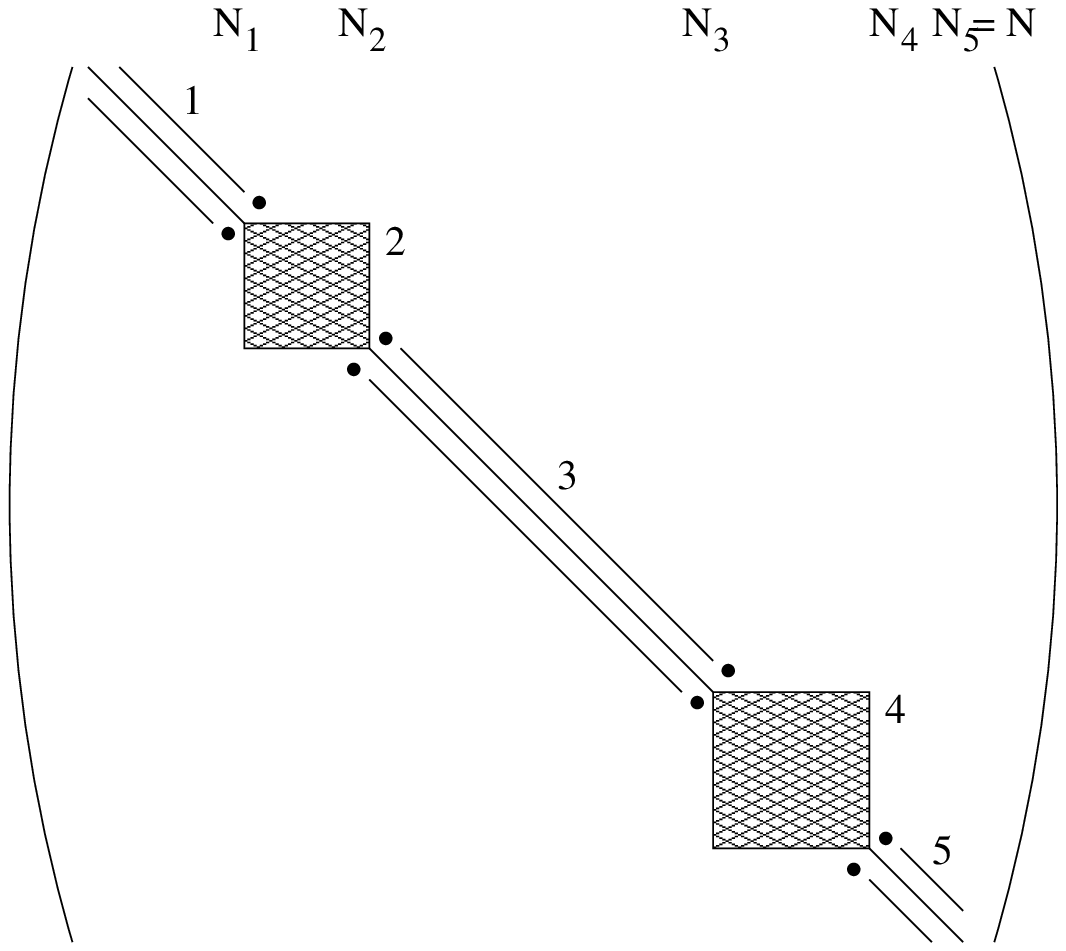}
\end{center}
\caption[]{Structure of the inverse Green function for stripe-like leads: block 1,3 and 5 are tridiagonal 
(left, right overhang and middle part of wire), block 2 and 4 are full matrices (left and right 
contact region), the dots represent the connecting elements; $(N_j,N_j)$ is the lower right 
element of the $j$-th block} \label{strucgreen}
\end{figure}

In order to compute the right-hand side of the flow equations (\ref{floweq}) and (\ref{floweq2}) for 
the self-energy, one needs the tridiagonal elements of the Green function $\mathcal{G}$. In the case 
of 1d leads $\mathcal{G}^{-1}$ itself is tridiagonal and these elements can be computed in 
order $N$ time~\cite{AndergfRG,TriInv} with $N$ being the length of the wire.
In the geometry with stripe-like leads, introduced in section \ref{GenRel}, the inverse Green function is 
of the structure shown in figure \ref{strucgreen}. Large tridiagonal blocks (the middle and overhanging 
parts of the wire) are connected on the sub- and 
superdiagonal to (small) full blocks (the contact regions).
We can exploit the tridiagonal structure of large parts of this matrix enabling us to treat lattice systems with 
up to $10^5$ sites\footnote{Our calculations were performed on a standard desktop PC. Thus the maximum system 
size reachable can be easily extended by using more elaborate hardware.}. To achieve this, we make multiple use of the projection 
formalism introduced in section \ref{GenRel} to split the problem such that we only have to 
invert matrices of size and structure of the \emph{isolated} blocks 1 to 5 (figure \ref{strucgreen}). 
This allows us to use the effective order $N$ time algorithm for the tridiagonal parts and
some standard algorithm~\cite{Schwarz} for the (small) full matrices. Thus, the bottleneck of 
this algorithm is the size of the full matrices.

In the first step of the algorithm the full Hamiltonian $h$ appearing in the Green 
function $\mathcal{G}(z)=\left(z-h\right)^{-1}$ is split into five smaller parts $h_j$, referring to 
block $j\in\{1,..,5\}$ respectively, and elements $h^{j,j+1}_{\rm con}=\beta_{j,j+1}|{N_j}\rangle\langle{N_j+1}|+{\rm H.c.}$ with $j\in\{1,..,4\}$ connecting block $j$ and $j+1$ (with $\beta_{j,j+1}\in\mathbb{R}$), i.e.
\begin{equation}\label{defofh}
h=\sum_{j=1}^5h_j+\sum_{j=1}^4h^{j,j+1}_{\rm con}
\end{equation}
We next define the projection operators $P_j$ on each block. Then we project the full Green function 
$\mathcal{G}$ onto the block, whose elements we want to compute, thereby reducing the effort of 
inverting the full matrix to the inversion of an effective matrix of size and 
structure of block $j$ of the inverse Green function. We will exemplify this by calculating 
the elements of block 1 as well as those elements on the sub- and superdiagonal connecting block 
1 and 2. Making subsequent use of the projection formula (\ref{profo}), we obtain
\begin{eqnarray*}
P_1&\mathcal{G}P_1=\\
&\left(z-h_1-\beta_{1,2}^2|{N_1}\rangle\langle{N_1+1}|\tilde{\mathcal{G}}^{(2)}|{N_1+1}\rangle\langle{N_1}|\right)^{-1},\\
P_2&\tilde{\mathcal{G}}^{(2)}P_2=\\
&\left(z-h_2-\beta_{2,3}^2|{N_2}\rangle\langle{N_2+1}|\tilde{\mathcal{G}}^{(3)}|{N_2+1}\rangle\langle{N_2}|\right)^{-1},\\
P_3&\tilde{\mathcal{G}}^{(3)}P_3=\\
&\left(z-h_3-\beta_{3,4}^2|{N_3}\rangle\langle{N_3+1}|\tilde{\mathcal{G}}^{(4)}|{N_3+1}\rangle\langle{N_3}|\right)^{-1},\\
P_4&\tilde{\mathcal{G}}^{(4)}P_4=\\
&\left(z-h_4-\beta_{4,5}^2|{N_4}\rangle\langle{N_4+1}|\tilde{\mathcal{G}}^{(5)}|{N_4+1}\rangle\langle{N_4}|\right)^{-1},\\
P_5&\tilde{\mathcal{G}}^{(5)}P_5=\left(z-h_5\right)^{-1}\\
\end{eqnarray*}
with $\tilde{\mathcal{G}}^{(i)}=\left(z-\sum_{j=i}^5h_j+\sum_{j=i}^4h^{j,j+1}_{\rm con}\right)^{-1}$. We
calculate the elements of block 1 of the Green function by going through this hierarchy
from bottom to top, having only to invert matrices of size and structure of the isolated 
blocks at each step. The computations for the other blocks of $\mathcal{G}$ can be performed similarly.

The element $\langle N_1+1|\mathcal{G}|N_1\rangle$ can be calculated by splitting $h$ 
into $\tilde{h}=\sum_{j=1}^5h_j+\sum_{j=2}^4h^{j,j+1}_{\rm con}$ and $h^{1,2}_{\rm con}$ to write
\begin{equation*}
{\mathcal G}=\tilde{\mathcal{G}}+\tilde{\mathcal{G}}h^{1,2}_{\rm con}
\tilde{\mathcal{G}}+\tilde{\mathcal{G}}h^{1,2}_{\rm con}\mathcal{G}h^{1,2}_{\rm con}\tilde{\mathcal{G}} 
\end{equation*}
with $\tilde{\mathcal{G}}=\left(z-\tilde{h}\right)^{-1}$. After projecting $\mathcal{G}$ onto the 
relevant block $P_1\mathcal{G}P_2$, one ends up with
\begin{eqnarray*}
\langle N_1+1|\mathcal{G}|N_1\rangle = \frac{\beta_{1,2}\langle N_1|\tilde{\mathcal{G}}|N_1\rangle
\langle N_1+1|\tilde{\mathcal{G}}|N_1+1\rangle}{1-\beta_{1,2}^4\langle N_1|
\tilde{\mathcal{G}}|N_1\rangle^2\langle N_1+1|\tilde{\mathcal{G}}|N_1+1\rangle^2}\\
\times\left(1+\beta_{1,2}^2\langle N_1|\tilde{\mathcal{G}}
|N_1\rangle\langle N_1+1|\tilde{\mathcal{G}}|N_1+1\rangle\right)
\quad.
\end{eqnarray*}
The relevant elements of $\tilde{\mathcal{G}}$ can be efficiently calculated using the algorithm 
introduced above for the blocks. Again the calculation of the other elements 
$\langle N_j+1|\mathcal{G}|N_j\rangle$ can be performed analogously.

For the calculation of the transmission probability (\ref{transampl}) we need all matrix elements 
$\langle j'|\mathcal{G}|j\rangle$ with $j'\in\{N_1+1,..,N_2\}$ and $j\in\{N_3+1,..,N_4\}$ of the Green 
function $\mathcal{G}^{-1}$. We use an algorithm which only requires the inversion of matrices 
of size and structure of the isolated blocks 1 to 5. We split the full Hamiltonian (\ref{defofh}) into blocks $\tilde{h}=\sum_{j=1}^5h_j$ and 
connecting elements $h_{\rm con}=\sum_{j=1}^4h^{j,j+1}_{\rm con}$ and decompose
\begin{equation*}
 \mathcal{G} = \tilde{\mathcal{G}}+\tilde{\mathcal{G}}h_{\rm con}\tilde{\mathcal{G}}
+\tilde{\mathcal{G}}h_{\rm con}\mathcal{G}h_{\rm con}\tilde{\mathcal{G}}
\end{equation*}
with $\tilde{\mathcal{G}}=\left(z-\tilde{h}\right)^{-1}$. Multiplication with projection operators $P_2$ and $P_4$ 
yields
\begin{eqnarray*}\label{mastereq}
P_2\mathcal{G}P_4 &= P_2\tilde{\mathcal{G}}\left(\beta_{2,3}|N_2\rangle\langle N_2+1|+\beta_{1,2}|N_1+1\rangle\langle N_1|\right)\\
& \mathcal{G}\left(\beta_{3,4}|N_3\rangle\langle N_3+1|+\beta_{4,5}|N_4+1\rangle\langle N_4|\right)\tilde{\mathcal{G}}P_2\quad.
\end{eqnarray*}
The elements of $\tilde{\mathcal{G}}$ can be easily computed as the disconnected blocks of $\tilde{\mathcal{G}}^{-1}$ can be inverted individually. The elements $\langle N_2+1|\mathcal{G}|N_3\rangle$, 
$\langle N_1|\mathcal{G}|N_3\rangle$, $\langle N_3+1|\mathcal{G}|N_4\rangle$, $\langle N_1|\mathcal{G}|N_4+1\rangle$
can be calculated by splitting $h$ into blocks and bonds and projecting onto the blocks needed. Details of the rather lengthy procedure can be found elsewhere~\cite{waechter}.
\end{appendix}

\end{document}